\documentclass[prd,aps,showpacs,showkeys]{revtex4-1}

\usepackage{bm}
\newcommand{\bb}{\begin{eqnarray}}
\newcommand{\ee}{\end{eqnarray}}

\begin{document}

\title{ \bf Zero-energy states of fermions in the field
of Aharonov--Bohm type in 2+1 dimensions}
\author{V.R. Khalilov}
\affiliation{Faculty of Physics, Moscow State University, 119991,
Moscow, Russia}

\begin{abstract}
The quantum-mechanical problem of constructing a
self-adjoint Hamiltonian for the Dirac equation in an
Aharonov--Bohm field in 2+1 dimensions is solved with taking into
account the fermion spin. The one-parameter family of self-adjoint extensions
is found for the above Dirac Hamiltonian with particle spin.
The correct domain of the self-adjoint
Hamiltonian extension selecting by means of acceptable
boundary conditions  can contain  regular and
singular (at the point ${\bf r}=0$) square-integrable functions
on the half-line with measure $rdr$. We argue that the physical
reason of the existence of  singular solutions is the additional
attractive potential, which appear
due to the interaction between the  spin magnetic moment of
fermion and Aharonov--Bohm magnetic field. For some range of parameters
there are bound fermionic states. It is shown that
fermion (particle and antiparticle)
states with zero energy are intersected what signals on
the instability of quantum system and the possibility of
a fermion-antifermion pair creation  by the static
external field.
\end{abstract}



\maketitle

\section{Introduction}

The quantum Aharonov--Bohm (AB) effect \cite{1} has been analyzed in various physical
situations in numerous works (see e.g., Ref. \cite{AB-review}).
When an electron travels in an Aharonov-Bohm field in which the
magnetic flux is restricted to a small-radius tube topologically
equivalent to a cylinder, the electron wave function acquires a
geometric phase. The AB vector potential can produce observable
effects because the relative (gauge invariant) phase of the
electron wave function, correlated with a nonvanishing  gauge
vector potential in the domain where the magnetic field vanishes,
depends on the total magnetic flux \cite{khu}.

When the external field configuration has the cylindrical
symmetry, a natural assumption is that the relevant quantum
mechanical system is invariant along the symmetry ($z$) axis and
the system then becomes essentially two-dimensional in the $xy$
plane. So, such models can be reduced to the (2+1)-dimensional ones. In
Refs. \cite{phgj,phg,ampw,aw} it was observed that solutions for
the Dirac equation in an Aharonov--Bohm field in 2+1 dimensions
are the Dirac equation solutions in infinite cosmic strings in 3+1
dimensions. Solutions to the two-component Dirac equation in the
AB potential were first obtained and discussed by Alford and
Wilczek in Ref. \cite{aw} in a study  of the interaction of cosmic
strings with matter. Relativistic quantum AB effect was studied in
Ref. \cite{hkh} for the free and bound fermion states by means of
exact analytic solutions of the Dirac equation in 2+1 dimensions
for a combination of an AB potential and the Lorentz three-vector
and scalar Coulomb potentials.  The effect of vacuum polarization
in the field of infinitesimally thin solenoid is recently
investigated in \cite{jmpt} and a wonderful phenomenon is
revealed: the induced current is finite periodical function of the
magnetic flux.

In \cite{ivt} it was observed that the Hamiltonian for the
Aharonov--Bohm problem is essentially singular and hence it cannot
be immediately defined in the domain $[0,\infty)$ for any
differentiable and enough rapidly decreasing (at $r\to \infty$) functions
in the Hilbert space of square-integrable functions
on the half-line with measure $rdr$. Usually, the Hamiltonians are
symmetric operators in its natural domain. The problem of
constructing a self-adjoint Hamiltonian is to find all
self-adjoint extensions of given symmetric operator and then to
select correct self-adjoint extension. The correctness of the
known Aharonov--Bohm solutions for scattering problem was analyzed
for spinless particles in \cite{ivt}, in which a self-adjoint extension of the
Hamiltonian is selected by physical condition -``the principle of
minimal singularity'': the Hilbert-space functions for which the
Hamiltonian is defined must not be singular.

A one-parameter self-adjoint extension of the Dirac Hamiltonian in
2+1 dimensions in the pure AB field  was constructed by means of
acceptable boundary conditions in \cite{phg,ampw,phgj}. In
\cite{phg} it was shown that the domain of the self-adjoint
Hamiltonian extension can contain, together with regular,
square-integrable functions on the half-line with measure $rdr$
and singular at $r=0$ as well as it was constructed a formal
solution, which describes a bound fermion state in the field of
cosmic string. 

Note that the usual four-component Dirac equation in 2+1
dimensions (in the absence of $z$ coordinate) decouples into two
uncoupled two-component Dirac equations for spin projection $s=+1$
and $s=-1$. Thus, the two-component Dirac  equation describes the
planar motion of relativistic electron having only one projection
of three-dimensional spin vector. The upper (``large'') and lower
(``small'') components of the two-component wave function are
interpreted in terms of positive- and negative-energy solutions of
the Dirac equation in 2+1 dimensions. The particle spin in the
two-component Dirac equation was artificially introduced in
\cite{crh} as a new parameter. The term including this new
parameter appears in the form of an additional delta-function
interaction of spin with magnetic field in the Dirac equation
squared.

 In this paper we would like to study  how the fermion spin affects
the properties of a bound Dirac fermion in an Aharonov-Bohm field
in 2+1 dimensions. We find the wave function of bound states,
derive an equation implicitly determining the fermion energy and
study the behavior of relativistic energy levels of spin-one-half
fermion in the AB field. It is shown  that the lowest energy levels of particles and
antiparticles intersect upon adiabatic variation of the magnetic
flux $\Phi$ between two integers $n$, i.e. when $\Phi=\Phi_0
(n+1/2)$, where $\Phi_0\equiv 2\pi/|e|$ is the elementary magnetic
flux  and $e$ is the fermion (electron)
charge.

The spectrum of Dirac's equation under consideration
is symmetric with respect to the change of the sign of energy and
the states with precisely zero energy exist.
Jackiw and Rebbi \cite{jacr} were observed
that, in a charge conjugation symmetric theory of one-dimensional
Dirac fermions interacting with a solitonic background field (the
kink), the vacuum acquires a fractional fermion charge $\pm 1/2$ due to
the existence of fermion states with zero energy (zero modes).
The states with zero energy exactly in the middle of spectrum have
been known to exist when the mass-term forms a vortex in the configurational
space \cite{jacro}. The problem of zero-energy states of the two-dimensional
Dirac Hamiltonian with a unit vortex in the mass-term is considered
in the presence of pseudo magnetic field in
the context of fractionalization by Jackiw and Pi in \cite{jacpi}
and in the presence of pseudo as well as true magnetic field in \cite{her}.

In the presence of a vector potential, the Dirac Hamiltonian does
not exhibit a charge conjugation symmetry since a charge coupling
treats particles and antiparticles differently. So the existence
of fermion states with zero energy does not
necessarily imply a fractional fermion number \cite{hokhl} but
the intersection of energy levels signals on
the instability of quantum system.

It is well to note that the possibility of existence of weakly
bound electron states was shown in \cite{KhaHo07,kh1} due to the
interaction between the three-dimensional spin magnetic moment of
electron and magnetic field of infinitely thin solenoid with
applying solutions of the Pauli equation in 3+1 dimensions.
We shall adopt the units where $c=\hbar=1$.

\section{Solutions to the Dirac equation in 2+1 dimensions in
an Aharonov-Bohm field for the scattering problem}

The Dirac equation for a fermion of  mass $m$ and charge $e>0$ in
2+1 dimensions  in the potential $A_{\mu}$ is
 \bb
 (\gamma^{\mu}
P_{\mu} - m)\Psi = 0, \label{Dirac}
 \ee
 Here the Dirac $\gamma^{\mu}$ matrices are conveniently defined in
terms of the Pauli spin matrices as (see, \cite{crh})
 \bb
\gamma^0= \sigma_3,\quad \gamma^1=is\sigma_1,\quad
\gamma^2=i\sigma_2, \label{1spin}
 \ee
 and $s$ is a new parameter
characterizing twice the spin value $s=\pm 1$ for spin ``up" and
``down", respectively, (see, \cite{crh}), $\hat P_{\mu} =
-i\partial_{\mu} - eA_{\mu}$ is the generalized electron momentum
operator.

We seek solutions of Eq. (\ref{Dirac}) in an Aharonov--Bohm field
 \bb
A^0=0,\quad A_r=0,\quad A_{\varphi}=\frac{B}{r}, \quad
r=\sqrt{x^2+y^2}, \quad \varphi=\arctan(y/x) \label{eight}
 \ee
in the form
  \bb
 \Psi(t,{\bf x}) = \frac{1}{\sqrt{2\pi}}\exp(-iEt+il\varphi)
\psi(r, \varphi)~, \label{three}
 \ee
where $E$ is the electron energy, $l$ is an integer, and
$\psi_l(r, \varphi)$ is a two-component function ({\it i.e.} a
$2$-spinor)
 \bb
  \psi(r,
\varphi) = \left( \begin{array}{c}
f_1(r)\\
f_2(r)e^{is\varphi}
\end{array}\right).
\label{four}
 \ee
The wave function $\Psi$ is an eigenfunction of the conserved
total angular momentum $J_z\equiv L_z+ s\sigma_3/2$, where
$L_z\equiv -i\partial/\partial\varphi$ with eigenvalue $j=l+s/2$.

It is seen that the radial Hamiltonian $h_r$ is singular at point
$r=0$ in the Aharonov--Bohm field and it cannot be immediately
defined for the class of functions to be self-adjoint operator.
So, we need to solve the eigenvalue problem for this operator,
which is \bb h_r\left(
\begin{array}{c}
f_1(r)\\
f_2(r)\end{array}\right)= \left[ \begin{array}{cc} m& sdf/dr+(l+\mu+s)/r \\
-sdf/dr+(l+\mu)/r& -m
\end{array}\right]\left( \begin{array}{c} f_1(r)\\
f_2(r)\end{array}\right) = E\left( \begin{array}{c} f_1(r)\\
f_2(r)\end{array}\right),\label{radh}\ee where $\mu\equiv eB$.

Because of the existence of finite magnetic flux inside solenoid
$\Phi=2\pi B$ the term including the spin parameter appears in the
form of an additional delta-function interaction of spin with
magnetic field of solenoid
 \bb
 {\bf H}=(0,\,0,\,H)=\nabla\times {\bf A}= B\pi\delta({\bf r})
\label{e1s} \ee in the Dirac equation (\ref{radh}) squared.  Note
that if $\mu$ is an integer $n$, then the magnetic field flux is
quantized as $\Phi=\Phi_0 n$. For the cosmic strings considered in
\cite{aw}, $\Phi=e/Q$, where $Q$ is the Higgs charge. The
additional (spin) potential  \bb -seB\frac{\delta(r)}{r}
\label{spterm} \ee in the Dirac equation (\ref{radh}) squared will
be taken into account by boundary conditions.

If $\mu$ is nonintegral then the regular solutions at $r=0$
of Eq. (\ref{radh}) for $E^2>m^2$ are
 \bb \left(
\begin{array}{c}
f_1(r)\\
f_2(r)\end{array}\right)=\frac{1}{N} \left(
\begin{array}{c}
\sqrt{E+m}J_{\nu}(pr)\\
\sqrt{E-m}J_{\nu+s}(pr)
\end{array}\right).
\label{sol1} \ee Here  $N$ is a normalization factor, $\nu =|l+\mu|$, $p =
\sqrt{E^2 - m^2}$, $\nu+s>0$ and $J_{\nu}(pr)$ is the regular Bessel function.
Singular Bessel functions at $r=0$ but square-integrable on the
half-line with measure $rdr$ also are admissible
quantum-mechanical solutions of Eq. (\ref{radh}). It is seen from Eq. (\ref{spterm})
that singular solutions (localized, obviously,
at the origin) can appear if only additional potential (\ref{spterm})
is attractive. Besides, rejecting singular solutions leads to
a loss of completeness in the angular basis \cite{phg,ktmp1}.

Since the spin term is invariant with respect to transformations
$e\to-e$, $s\to-s$, we can consider the case $e>0$ only. Let
$\mu>0$, then potential (\ref{spterm}) is attractive for $s=1$ and
repulsive for $s=-1$. So, singular solutions (localized  at $r=0$)
can appear if $\mu>0,\quad B>0,\quad s=1$ (particle state) or
$\mu<0,\quad B>0,\quad s=-1$ (antiparticle state). Written  \bb
\mu=[\mu]+\gamma\equiv n+\gamma, \label{divi}\ee where
$[\mu]\equiv n$ denotes the largest integer $\le \mu$, and \bb
1>\gamma> 0, \label{nonint}\ee we can easily find, that singular,
square-integrable functions are the Bessel functions of the order
$\gamma-1$ (upper spinor component) and $-\gamma$ (lower spinor
component) with $l=-n-1,\quad n\ge 0$.

\section{Self-adjoint extensions for the radial Dirac Hamiltonian}

We must construct the self-adjoint extensions of the radial Dirac
Hamiltonian $h_r$ and, then, select a needed extension by means of
some physical condition. The problem is solved for symmetric
operators  by the method of deficiency indices developed by von
Neumann (see, for example, \cite{nai,crein,reed}). In our problem
$h_r$ is symmetric operator if, for arbitrary spinors $f(r)$ and
$g(r)$, \bb
 \int\limits_{0}^{\infty}g^{\dagger}(r)h_r f(r)r dr =
 \int\limits_{0}^{\infty}[h_rg(r)]^{\dagger}f(r)r dr,  \label{sym}
 \ee
what leads to the following boundary condition \cite{phg}) \bb
\lim_{r\to 0} rg^{\dagger}(r)i\sigma_2 f(r)=0. \label{bounsym}
 \ee

A symmetric operator $h$ is self-adjoint, if its domain $D(h)$
coincides with the domain of its adjoint operator. Since the
defect subspace contains functions, which are singular at $r=0$,
the adjoint operator has a larger domain. So, it is natural to
posit the boundary condition (\ref{bounsym}) in the defect
subspace.

Let  the radial Dirac Hamiltonian $h_r$ (\ref{radh}) has the
domain $D(f)$, where $f(r)$ is absolutely continuous functions,
square integrable  on the half-line $[0,\infty)$ with measure
$rdr$ and regular at $r=0$. We must construct the defect subspaces
$D^{\pm}$ of adjoint operator $h^{\dagger}_r$ with eigenvalue $\pm
im$ ($m$ is the fermion mass) \bb
h^{\dagger}_rf(r)\equiv \left[ \begin{array}{cc} m& sdf/dr+(l+\mu+s)/r \\
-sdf/dr+(l+\mu)/r& -m
\end{array}\right]\left(
\begin{array}{c}
f^{\pm}_1(r)\\
f^{\pm}_2(r)\end{array}\right)=\pm i m \left( \begin{array}{c} f^{\pm}_1(r)\\
f^{\pm}_2(r)\end{array}\right).\label{radhdag}\ee

Then, any self-adjoint extension of $h_r$ can be constructed by
the isometries $D^+\to D^-$. This extension can be fixed by a
parameter $\theta$: \bb f^+(r)\to e^{i\theta}f^-(r).
\label{param}\ee The correct domain for the self-adjoint extension
$h^{\theta}$ of $h_r$ is given by  \bb D(h^{\theta})= D(h_r) +
C[f^+(r)+ e^{i\theta}f^-(r)],\label{cordom}\ee where $C$ is
arbitrary complex constant, and $\theta$ is arbitrary parameter
but fixed for given extension ($2\pi>\theta>0$).

In our case, it is simpler to use solutions of Eq.
(\ref{radhdag}), with $m=0$ in the body of operator
$h^{\dagger}_r$: \bb \left(
\begin{array}{c}
f^{\pm}_1(r)\\
f^{\pm}_2(r)\end{array}\right)=N\left(
\begin{array}{c}
K_{\nu}(mr)\\
se^{\mp i\pi/2}K_{\nu+s}(mr)
\end{array}\right),
\label{soldag} \ee where $N$ is a normalization factor, and
$K_{\nu}$ is the MacDonald function. Singular, square integrable
functions are the MacDonald functions of the order $\gamma-1$ (upper
spinor component) and $\gamma$ (lower spinor component) with
$l=-n-1,\quad n\ge 0$. Parameterized spinors of the defect
subspaces in the correct domain of self-adjoint extension
(\ref{cordom}) can be easily constructed in the simple form  \bb
2Ce^{i\theta/2}\left(
\begin{array}{c}
K_{\gamma-1}(mr)\cos(\theta/2)\\
-sK_{\gamma}(mr)\sin(\theta/2)
\end{array}\right).
\label{solcor} \ee

Taking into account formula $K_{\nu}(z)=K_{-\nu}(z)$ and
asymptotic behavior $K_{\nu}(x)\cong 2^{\nu-1}\Gamma(\nu)/x^{\nu}$
at $x\to 0$, where $\Gamma(\nu)$ is the  gamma function, we can
easily find that boundary condition (\ref{bounsym}) will be
satisfied for arbitrary spinor with its asymptotic behavior  \bb
\lim_{mr\to 0} f(mr)\sim \left(
\begin{array}{c}
(mr)^{\gamma-1}\sin(\theta^*/2)\\
-s(mr)^{-\gamma}\cos(\theta^*/2)
\end{array}\right).
\label{solasymp} \ee
Here
 \bb
\tan(\theta^*/2)=\frac{\Gamma(1-\gamma)}{2\Gamma(\gamma)}\tan(\theta/2)
\label{connect} \ee
and $2\pi>\theta^*>0$.

\section{Bound fermion states in the field of a cosmic string}

For nonintegral $\mu$ there exists a formal solution of Eq.
(\ref{radh}) with $m^2>E^2$ that describes a bound fermion state
with the spin $s$ \bb \left(
\begin{array}{c}
f_1(r)\\
f_2(r)\end{array}\right)=N\left(
\begin{array}{c}
\sqrt{m+E}K_{\gamma-1}(kr)\\
s\sqrt{m-E}K_{\gamma}(kr)
\end{array}\right),
\label{solboun} \ee where  $N$ is a normalization factor, $k =
\sqrt{m^2 - E^2}$. This bound fermion state
exists for the range  $2\pi>\theta^*>\pi$. The appearance of bound state
``suggests that this range of parameters in the effective
Hamiltonian parameterizes nontrivial physics in the core'' of
cosmic string \cite{phg}. We see that the physical reason for the
existence of bound state can be the appearance of
additional attractive (for instance, $D\delta({\bf r})$ type)
potential in the core of cosmic string.

From (\ref{solasymp}) and (\ref{solboun}) we derive an equation,
which implicitly determines the bound state energies of particle
($m\le E\le 0$) and antiparticle ($0\le E\le -m$) in the form \bb
\frac{(m+E)^{\gamma}}{(m-E)^{1-\gamma}}=-(2m)^{2\gamma-1}
\frac{\Gamma(\gamma)}{\Gamma(1-\gamma)}\tan\frac{\theta^*}{2}.
\label{ener}\ee Indeed, from the continuity consideration, we can
conclude that particle states are the states that tend to the
boundary of the continuous spectrum $E=m$ upon infinitely slow
switching off the external field; under such switching off
antiparticle states tend to the boundary $E=-m$ (see, e.g.,
\cite{blp}). So Eq. (\ref{ener}) as a function of $\gamma$
describes two energy (particle and antiparticle) curves.

For $\theta^*=3\pi/2$ these curves are symmetric with respect to the
horizontal line $E=0$ and when parameter $\gamma$ changes from $0$
to $1/2$, Eq. (\ref{ener}) determines the energy of bound particle
state in the region $m\le E\le 0$. Denoting  $x=E/m$  Eq.
(\ref{ener}) can be written for each curve in the region
$0<\gamma<1/2$ as \bb \pm x=b(1-x^2)^{1-\gamma}-1,\quad
b=2^{2\gamma-1}\pi^{-1}\Gamma^2(\gamma)\sin\pi\gamma>0.
\label{ener1}\ee Two curves intersect the horizontal line $E=0$ at
$\gamma=1/2$.

We see that in the AB field, indeed the Dirac Hamiltonian in 2+1
dimensions does not exhibit a charge conjugation  but
fermion states with zero energy exist.
For $\gamma=1/2,\quad s=1$ the particle wave function (zero mode)
in the lowest energy state is \bb \left(
\begin{array}{c}
f_1(r)\\
f_2(r)\end{array}\right)=N\left(
\begin{array}{c}
K_{1/2}(mr)\\
K_{1/2}(mr)
\end{array}\right).
\label{groundsol} \ee The antiparticle wave function with $E=0$
for $\gamma=1/2,\quad s=-1$ can be obtained by means
 the charge conjugation operator, which, in our case, is
 $C=i\sigma_2$. Hence, the antiparticle wave function is
\bb \left(
\begin{array}{c}
f_1(r)\\
f_2(r)\end{array}\right)=N\left(
\begin{array}{c}
K_{1/2}(mr)\\
-K_{1/2}(mr)
\end{array}\right).
\label{groundant} \ee

Thus, when the parameter $\gamma$ changes adiabatically from $0$ to $1/2$,
(the magnetic flux $\Phi$ changes from $\Phi=2\pi n/|e|$
to $\Phi=2\pi(n+1/2)/|e|$) the energy split between bound states of particle and
antiparticle vanishes and their lowest energy levels
intersect. The intersection of energy levels signals on
the instability of quantum system, i.e. the vacuum,  and the possibility of creation of a
fermion-antifermion pair from the vacuum by the static
external field. The latter is not valid for the electron-positron pair production
by the real Aharonov--Bohm field of thin solenoid since the electron energy
levels in the AB field has to be defined from the corresponding Dirac
equation in 3+1 dimensions. But solutions of the Dirac equation in
an Aharonov--Bohm field in 2+1 dimensions describe relativistic
fermions in the field of cosmic string in 3+1 dimensions, so, one
can hope that obtained results will be helpful in studying the
behavior of fermions in the field of cosmic string.

It is helpful to compare the fermion's creation in the AB field with
the positron creation from the vacuum by a strong Coulomb potential field
$U(r)=-a/r, a>0$ in the quantum electrodynamics (QED). When $a$ changes adiabatically
to ``the critical charge'' $a_{cr}$ the lowest level of electron (with charge $e<0$)
tends to the boundary of the lower continuum of energies. For $a>a_{cr}$ it intersects
the above boundary and the vacuum becomes unstable, which leads to
the appearance of a bound (vacuum) state in the lower continuum and results in the
positron production in a free state (see, for example, \cite{gmamo}).
Due to the existence of bound state the vacuum simultaneously
can acquire  negative electric charge \cite{gmamo,khl2}. In the AB field
in 2+1 dimensions, when $\gamma$ changes adiabatically to $1/2$,
the zero-energy (fermion and antifermion) states simultaneously
exist and the vacuum, evidently, can acquire a magnetic moment equal
 to the two spin magnetic moment of the fermion.

 It should be noted that the existence of induced charge density,
which can change the critical charge, in a strong Coulomb
field due to vacuum polarization does not significantly
change results concerning to the positron production from the vacuum by a Coulomb
field (see, \cite{gmamo}); the induced charge density in the field of infinitesimally
thin solenoid due to vacuum polarization  is exactly equal to zero \cite{jmpt}. Of course,
all the discussed  questions require more subtle study since the effects at
$\gamma\ge 1/2$ are of a many-particle nature and to describe
them, we need a quantum field theory formalism.

\begin{acknowledgments}

\vskip 0.5cm

This paper was supported, in part, by the Program for Leading
Russian Scientific Schools (Grant No. NSh-3312.2008.2).

\end{acknowledgments}  

\end{document}